\documentclass[preprint,showpacs,preprintnumbers,amsmath,amssymb]{revtex4}

\usepackage{graphicx}
\usepackage{dcolumn}
\usepackage{bm}

\begin{document}

\preprint{cond-mat/0210630}

\title{Hard-sphere limit of soft-sphere model for 
granular materials:
Stiffness dependence of steady granular flow}

\author{Namiko Mitarai}
 \email{namiko@stat.phys.kyushu-u.ac.jp}
\author{Hiizu Nakanishi}%
 \email{naka4scp@mbox.nc.kyushu-u.ac.jp}
\affiliation{%
Department of Physics, Kyushu University 33,
Fukuoka 812-8581, Japan
}%

\date{\today}

\begin{abstract}
Dynamical behavior of steady granular flow is investigated numerically in
the inelastic hard sphere limit of the soft sphere model.  We find 
distinctively different limiting behaviors for the two flow regimes,
i.e., the collisional flow and the frictional flow.  In the collisional
flow, the hard sphere limit is straightforward;
the number of collisions per particle per unit time converges
to a finite value and the total contact time fraction with other
particles goes to zero.  
For the frictional flow, however,
we demonstrate that the collision rate
diverges as the power of the particle stiffness so that the
time fraction of the multiple contacts remains finite even in the hard
sphere limit although the contact time fraction for the binary
collisions tends to zero.
\end{abstract}

\maketitle

\section{Introduction}
The interactions between grains in flowing granular material
are roughly classified into two types;
the impulsive contact (collision) 
with the momentum exchange and 
the sustained contact 
with the transmission of forces \cite{col-fri}.
The flow in which the impulsive contact is dominant
is called {\it collisional flow},
while the flow where the sustained contact 
dominates is called {\it frictional flow}.
These two types of flow may be found
in a simple geometry
such as granular flow on a slope.
The grains stay at rest 
when the inclination angle is too small.
If the inclination exceeds a critical angle,
the material starts flowing frictionally at first.
The flow becomes collisional when the inclination is large enough.

As for the collisional flow of granular material,
its dynamics has some analogy with molecular fluid,
and the kinetic theories based on inelastic binary collisions
of particles hold to some extent \cite{kinetic}.
On the other hand, the frictional flow is drastically different from
the molecular fluid, and we have little understanding on it.
Many models have been proposed for dense granular flows:
For example, some models take into account
the effect of non-local force transmission 
which comes from the network of contacting grains \cite{densemodel}.
In the experiments, however, it is difficult 
to specify the nature of sustained contact
in the dense flow.

For the simulations of granular dynamics, the following two models have
been commonly used, i.e.  the inelastic hard sphere model and the soft
sphere model.

In the inelastic hard sphere model, the particles
are rigid and the collisions are thus instantaneous, 
therefore
its dynamics can be defined through a few parameters that characterize
a binary collision because there are no many-body collisions \cite{duran}.  
The model is simple and there are very efficient algorithms to simulate 
it \cite{I99}, but
it describes basically only the collisional flow \cite{L94} 
because the sustained
contact is not allowed.  It is also known that the system often
encounters what is called the inelastic 
collapse \cite{K99,BM90,MY92,2Dcollapse}; 
the infinite number
of collisions take place among a small number of particles in a
finite time, thus the dynamics cannot be continued beyond that point
without additional assumption.

On the other hand, in the soft sphere model, that is sometimes called
the discrete element method (DEM) 
in the granular community \cite{duran,CS79}, 
the particles overlap during collision
and the dynamics is defined through the forces acting on the colliding
particles.  Collision takes finite time, and not only binary collision
but also many-body collision and sustained contact between particles
are possible, therefore, both the collisional and the frictional flows
may realize in the model.
Many researches have been done on granular flow down 
a slope using the DEM in both the collisional regime 
and the frictional regime \cite{P93,MN01,denseDEM}.

In actual simulations, however, the stiffness constants used in the
soft sphere model are usually much smaller than the one appropriate
for real material such as steel or glass ball \cite{YASU95}
because of numerical difficulty.
Therefore, some
part of sustained contact in simulations may be decomposed into binary
collisions if stiffer particles are used.  It is also not clear that
the frictional force in the sustained contact may be described by the same
forces with the one suitable for the collisional events.

It is, therefore, important to examine how the system behavior may
change as the stiffness constant increases in the soft sphere model,
or in other words, how the soft sphere model converges to the
inelastic hard sphere model in the infinite stiffness limit.
In this paper, we present the results of numerical simulations of the
granular flow using the soft sphere model, and investigate the system
behavior when we change the stiffness constant systematically with
keeping the resulting restitution constant unchanged.

In Sec. \ref{sec:2}, we briefly summarize an inelastic
hard sphere model that is used for collisional granular flows.
Then we introduce a simple soft sphere model 
for granular material and discuss how we take the
hard sphere limit 
keeping the restitution coefficient constant. 
In Sec. \ref{sec:3}, the simulation results are shown.
At first, 
the stiffness dependence of the steady state of a single particle rolling
down a slope is presented
to see how the inelastic collapse appears in the hard sphere limit.
The collisional flow and the frictional flow are examined.
We find that the interactions between particles 
in the collisional flow smoothly converge 
to binary collisions of inelastic hard spheres,
while those in frictional flow shows non-trivial behavior;
the behavior is also different from the ``inelastic collapse'' 
in the single particle system.
The summary and the discussion are given in Sec. \ref{sec:4}.

\section{Hard sphere model and soft sphere model}
\label{sec:2}
In this section, we introduce the hard sphere model
and the soft sphere model for granular material.
For simplicity, we consider a two dimensional system
and grains are modeled by two dimensional disks.

After showing the correspondence of parameters
in the soft sphere model to those of the hard sphere model, 
we briefly summarize the phenomenon 
called inelastic collapse, which 
is the singular behavior in the inelastic hard sphere system. 
At the last part of this section, we discuss the limiting behavior
of the soft sphere model that corresponds to the
inelastic collapse of the hard sphere model.

\subsection{Inelastic hard sphere model}\label{hard}
In hard sphere models,
collisions between particles or between a particle and
a wall are considered to happen instantaneously,
and its dynamics is defined by the binary collision rule.
We consider the collision rule in terms of the normal and tangential
restitution coefficients and the sliding friction \cite{L94,L95}.

Let us
consider a collision between the two spheres $i$ and $j$
of the diameters $\sigma_i$ and $\sigma_j$
and the masses $m_i$ and $m_j$ 
at the contact positions $\bm{r}_{i}$ and $\bm{r}_j$, 
respectively.
Prior to the collision, the disks have
velocities $\bm{c}_{i}$ and $\bm{c}_j$
and angular velocities $\bm \omega_{i}$  and $\bm \omega_j$.
Then the relative velocity of the point of contact $\bm v_{ij}$
is given by
\begin{equation}
\bm v_{ij}=(\bm c_i-\bm c_j)
+\bm n \times 
\left(\frac{\sigma_i}{2}\bm \omega_i
+\frac{\sigma_j}{2}\bm \omega_j\right),
\label{eq:vij}
\end{equation}
where the normal vector $\bm{n}=\bm{r}_{ij}/|\bm{r}_{ij}|=(n_x,n_y,0)$
with $\bm{r}_{ij}=\bm{r}_i-\bm{r}_j$.

If $\bm v_{ij}'$ denotes the post-collisional relative velocity,
the collision rule for normal direction is
\begin{equation}
(\bm n\cdot \bm v_{ij}')=-e(\bm n\cdot \bm v_{ij}),
\label{eq:restie}
\end{equation}
where $e$ is the normal restitution coefficient
with $0\le e\le 1$.

In the case without sliding, the collision rule 
in the tangential direction is given by
\begin{equation}
(\bm n\times \bm v_{ij}')=-\beta(\bm n\times \bm v_{ij}),
\label{eq:betadef}
\end{equation}
where $\beta$ is
the tangential restitution coefficient 
with $-1\le\beta\le1$.
The sliding is taken into account so that
the tangential component of impulse does not exceed 
$\mu|\bm n\cdot \bm J|$ with 
the Coulomb friction coefficient $\mu$, 
where $\bm J$ is the momentum change of the particle $i$
through the collision.
Namely, when the momentum change of the particle $i$
through the collision rules
(\ref{eq:restie}) and (\ref{eq:betadef}) is $\bm{J}^{nos}$,
then $(\bm n\times \bm v_{ij}')$ is determined 
so that
\begin{equation}
|\bm n\times \bm J|=
\mbox{min}(\mu|\bm n\cdot \bm J|, |\bm n\times \bm{J}^{nos}|)
\end{equation}
is satisfied.
The detailed description of the rule is given in Refs. \cite{L94,L95}.

\subsection{Soft sphere model: DEM}
The DEM, or the soft sphere model \cite{CS79,duran}, is often
used to simulate the dynamics of granular materials. 
In the present work, we 
adopt the two dimensional one with 
the linear elastic force and dissipation.
When the two disks $i$ and $j$ at positions $\bm{r}_{i}$ and $\bm{r}_j$ 
with velocities $\bm{c}_{i}$ and $\bm{c}_j$
and angular velocities $\bm \omega_{i}$  and $\bm \omega_j$ are
in contact, the force acting on the particle $i$ from the particle $j$
is calculated as follows:
The normal velocity $v_n$, the tangential velocity $v_t$,
and the tangential displacement $u_t$ are
given by
\begin{eqnarray}
v_n=\bm n \cdot \bm{v}_{ij},&&
v_t=\bm t \cdot \bm{v}_{ij},\\
u_t&=&\int^t_{t_0}v_t\mbox{d}t,
\end{eqnarray}
where $\bm v_{ij}$
is given in Eq. (\ref{eq:vij}), and 
$\bm t=(-n_y,n_x,0)$.
Here, $t_0$ is the time when the two particles start to be in contact.
Then the normal force $F^n_{ij}$ 
and the tangential force $F^t_{ij}$ 
acting on the particle $i$ from the particle $j$ are given by
\begin{eqnarray}
F^n_{ij}&=&2M k_n\left(\frac{\sigma_i+\sigma_j}{2}
-|\bm{r}_{ij}|\right)-2M \eta_n v_n,
\label{eq:fn}\\
F^t_{ij}&=&\min(|h_t|,\mu|F_n|) \mbox{sign}(h_t)
\label{eq:ft}
\end{eqnarray}
with 
\begin{equation}
h_t=-2M k_tu_t-2M \eta_tv_t,
\end{equation}
where $k_{n}$ and $k_t$ are the elastic constants,
$\eta_{n}$ and $\eta_t$ are the damping parameters, 
$\mu$ is the Coulomb friction coefficient for sliding,
and $M=m_i m_j/(m_i+m_j)$ is the reduced mass.

\subsection{Hard sphere limit of soft sphere model}
When we adopt the linear dependence of the elastic and viscous force
on the overlap and the normal relative velocity as in Eq. (\ref{eq:fn}),
we can calculate the duration of contact $\tau_c$ and the
restitution coefficient for a normal binary collision $e$ \cite{SDW96};
they are given by
\begin{equation}
\tau _c=\frac{\pi}{\sqrt{2k_n-\eta_n^2}}
\label{eq:tauc}
\end{equation}
and 
\begin{equation}
e=\exp\left(-\frac{\pi\eta_n}{\sqrt{2k_n-\eta_n^2}}\right),
\label{eq:en}
\end{equation}
respectively.

Neglecting the sliding friction and the variation of ${\bm n}$
during the contact, the half period of 
the oscillation in the tangential displacement $u_t$
is estimated as \cite{SDW96}
\begin{equation}
\tau_s=\frac{\pi}{\sqrt{6k_t-9\eta_t^2}}
\end{equation}
for the two dimensional disks with 
the moment of inertia $I_i=m_i\sigma_i^2/8$.
Here we choose 
the parameters $k_t$ and $\eta_t$
so that the relation $\tau_s=\tau_c$, or,
\begin{equation}
\sqrt{6k_t-9\eta_t^2}=\sqrt{2k_n-\eta_n^2}
\label{eq:beta2}
\end{equation}
is satisfied.
Under this condition, the tangential restitution coefficient is given by
\begin{equation}
\beta=\exp(-3\eta_t \tau_c).
\label{eq:beta}
\end{equation}

Equations (\ref{eq:tauc}) and (\ref{eq:en}) can be rewritten as
\begin{eqnarray}
\eta_n&=&\left[\frac{2k_n}{(\pi/\ln e)^2+1}\right]^{1/2},
\label{eq:en2}\\
\tau_c&=&\left[\frac{\pi^2+(\ln e)^2}{2k_n}\right]^{1/2}.
\label{eq:tauc2}
\end{eqnarray}
Using Eqs. (\ref{eq:beta2})-(\ref{eq:tauc2}), 
we can take the inelastic hard sphere limit
of this model for given $e$ and $\beta$
by taking the $k_n\to \infty$ limit;
$\eta_n$, $\eta_t$, and $k_t$ are given by 
Eqs. (\ref{eq:en2}), (\ref{eq:beta}), and (\ref{eq:beta2}),
and the duration time of collision $\tau_c$ goes to zero
as Eq. (\ref{eq:tauc2}).

\subsection{Inelastic collapse}\label{sec:2.4}
It is well known that the inelastic hard sphere system
can undergo the {\it inelastic collapse}, i.e.,
the phenomenon where infinite collisions take place within
a finite period of time \cite{K99}.
The simplest example 
is the vertical bouncing 
motion of a ball under gravity,
but it has been shown that the inelastic collapse also occurs
in higher dimensional systems without 
gravity~\cite{K99,BM90,MY92,2Dcollapse}. 

The inelastic collapse never occurs
in the soft sphere model because of the finite length of the collision time.
However, it is worth to discuss 
what will happen in the soft sphere model in the simple situation
where the inelastic collapse occurs in the hard sphere model.
First, let us consider the vertical bouncing motion of a soft ball
under gravity. 
We assume the same force law between the ball and the floor
as in Eqs. (\ref{eq:fn}) and (\ref{eq:ft}),
except that we replace $2M$ by the mass of the ball. 
While the ball and the floor are in contact,
the equation of motion for the overlap of 
the ball and the floor, $z$, is given by
\begin{equation}
\ddot{z}+k_n z+\eta_n \dot{z}=g,
\end{equation}
with the acceleration of gravity $g$,
where the dots represent the time derivatives.
One can show that,
if the impact velocity $v_i$
is below a critical value $v_c$, which is $O(1/\sqrt{k_n})$ for 
large enough $k_n$, the ball stays in contact with the floor;
for $v_i \gg v_c$, 
the restitution coefficient $e_w$ 
can be considered as a constant.
Therefore, for a given initial 
impact velocity $v_0$, 
the number of necessary collisions $n_c$ for the ball to 
stay in contact with the floor is roughly
estimated by the condition $e_w^{n_c}v_0\sim v_c$, namely, 
$n_c\sim \ln\left(v_c/v_0\right)/\ln e_w$.
Because $v_c\sim 1/\sqrt{k_n}$ for large $k_n$, $n_c$ behaves as 
\begin{equation}
n_c \propto \ln k_n+\mbox{const.}
\label{eq:logk}
\end{equation}
in the hard sphere limit.
Thus $n_c$ diverges logarithmically when $k_n\to \infty$,
which corresponds to the inelastic collapse due to gravity.

In the case without gravity, the inelastic collapse
results in the ``many-body collision'' in the soft sphere model. 
For example, we consider three soft spheres 
in one dimension and assume that the binary collision can be approximated
by the collision law with a constant restitution coefficient.
In the situation where a particle goes back and forth 
between the two particles approaching each other,
the inelastic collapse may take place 
in the hard sphere model \cite{BM90};
three balls lose relative velocity completely 
in the limit of infinite collisions. 
In the soft sphere model, however,
when the interval between two collisions 
becomes smaller than the duration of contact $\tau_c$,
three balls are in contact at the same time,
namely, the three body collision occurs;
then they will fly apart. 
The number of collisions before the three body collision
will also diverge logarithmically in the hard sphere limit
because $\tau_c\propto 1/\sqrt{k_n}$.

On the other hand, one should note that
a many-body collision in the soft 
sphere model does not necessarily result
in the inelastic collapse in the hard sphere limit.
Actually, in most of the cases,
a many-body collision will be decomposed 
into a set of binary collisions in the hard sphere limit.

\section{Simulation Results}
\label{sec:3}
In this section, we investigate the stiffness
dependence of granular material on a 
slope in the following three situations:
(i) a single particle rolling down a slope,
(ii) the dilute collisional flow,
and (iii) the dense frictional flow.
We focus on the steady state in each situation and
compare the simulation data with changing 
$k_n$ systematically.
The particles are monodisperse in (ii), while they are
polydisperse in (iii) in order to avoid crystallization.

In the simulations, the parameters have been chosen to give
$e=\beta=0.7$, $\mu=0.5$ in the hard sphere limit.
All values are non-dimensionalized by the
length unit $\sigma$, the mass unit $m$, and the time unit
$\sqrt{\sigma/g}$.
Here, $\sigma$ is the diameter of the 
largest particle in the system 
and $m$ is the mass of that particle.
The second order Adams-Bashforth method 
and the trapezoidal rule are used 
to integrate the equations for the velocity and the position, 
respectively \cite{integ}.
Note that the time step for integration, $dt$, 
needs to be adjusted as $\tau_c$ becomes smaller.
All the data presented in the paper are 
results with $dt =\mbox{min}(\tau_c/100,10^{-4})$.
We have confirmed that the results do not change
for $dt\le \tau_c/100$ by calculating also
with $dt=\tau_c/50$ and $dt=\tau_c/200$ in the case of the single particle.

\subsection{A single particle rolling down a bumpy slope}\label{one}
\begin{figure}[thb]
\begin{center}
\includegraphics[width=7cm,angle=-7.47]{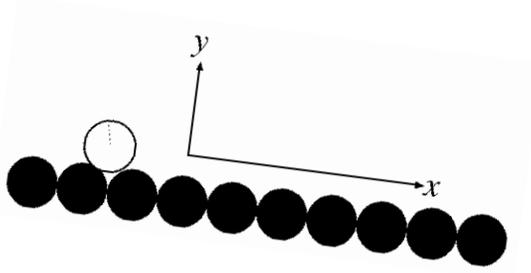}
\end{center}
\caption{A snapshot of a single 
ball rolling down a rough slope.}
\label{snap1}
\end{figure}

\begin{figure}[thb]
\begin{center}
\includegraphics[angle=-90,width=7cm]{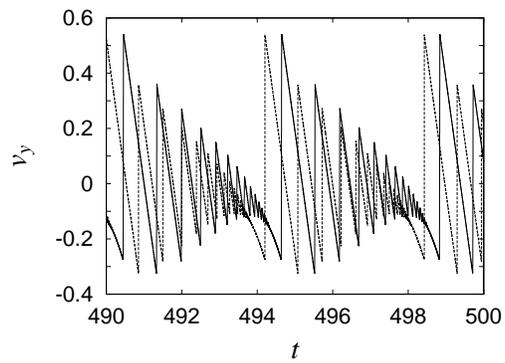}
\end{center}
\caption{Time evolution of $v_y$ with $k_n=2^{-1}\times 10^5$
(solid line) and $2^{17}\times 10^5$ (dashed line).}
\label{vy-evol}
\end{figure}

Let us first consider a single particle rolling down a bumpy slope.
It is known that 
the particle shows the steady motion
for a certain range of the inclination angle $\theta$ \cite{DBW96}.
In the simulations, we make the boundary rough
by attaching the same particles with the rolling one to the slope
with the spacing $0.002 \sigma$ (see Fig. \ref{snap1}).
For the chosen parameters with
the normal stiffness $k_n=2^{-1}\times 10^5$,
the range of $\theta$ for which steady motion is realized is
$0.11\lesssim\sin\theta\lesssim 0.14$ \cite{MN01}.
Here we fix the inclination angle to $\sin\theta=0.13$.
Figure  \ref{vy-evol} shows 
the time evolution of the velocity in 
the $y$ direction, $v_y$, 
with $k_n=2^{-1}\times 10^5$ (solid line)
and $k_n=2^{17}\times 10^5$ (dashed line). 
It is shown that $v_y$ behaves periodically;
this period ($\Delta t\sim 4$)
corresponds to the period for the particle 
to get past one particle at the floor.
The period hardly depends on the stiffness.

\begin{figure}[thb]
\begin{center}
\includegraphics[angle=-90,width=7cm]{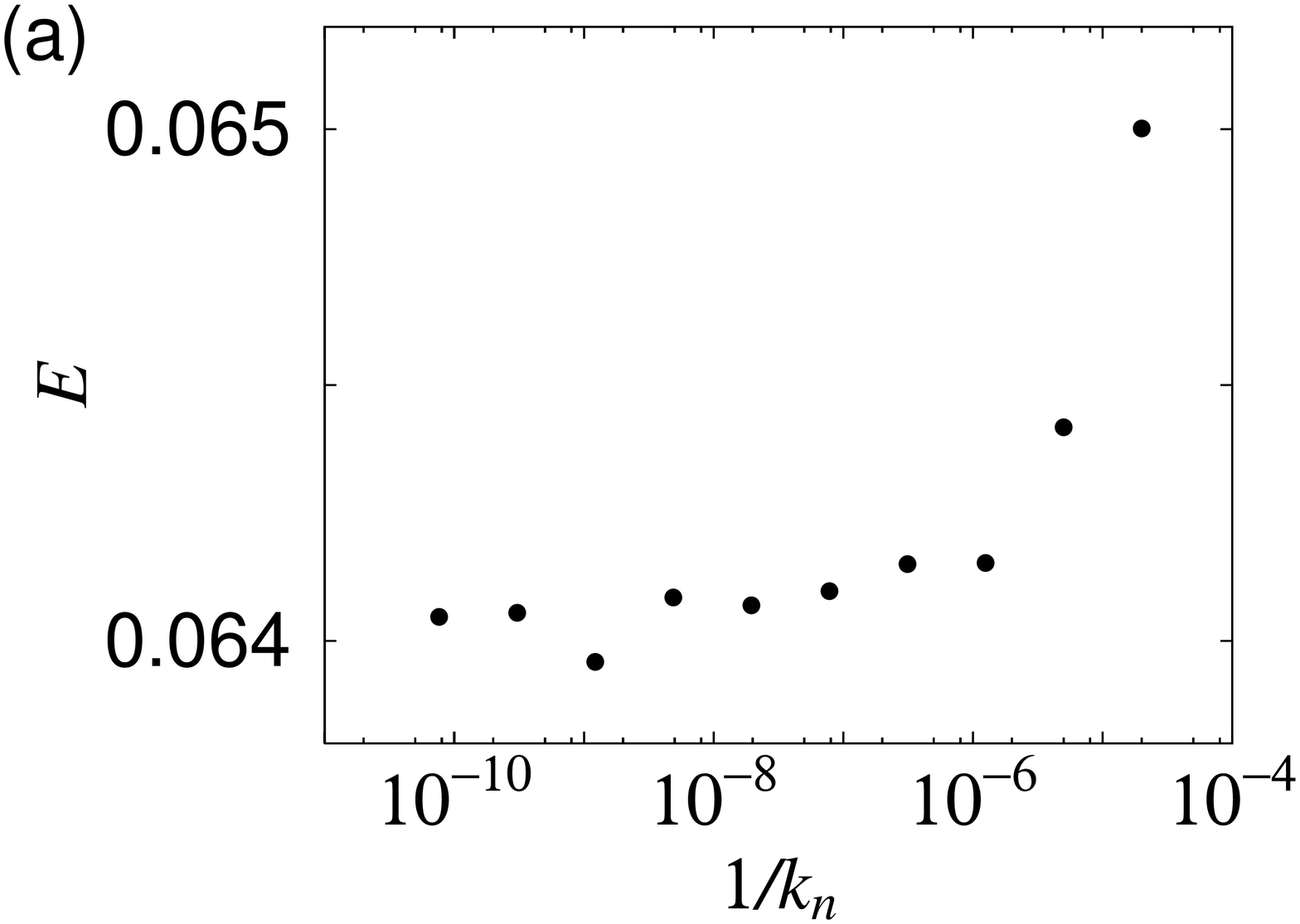}
\includegraphics[angle=-90,width=7cm]{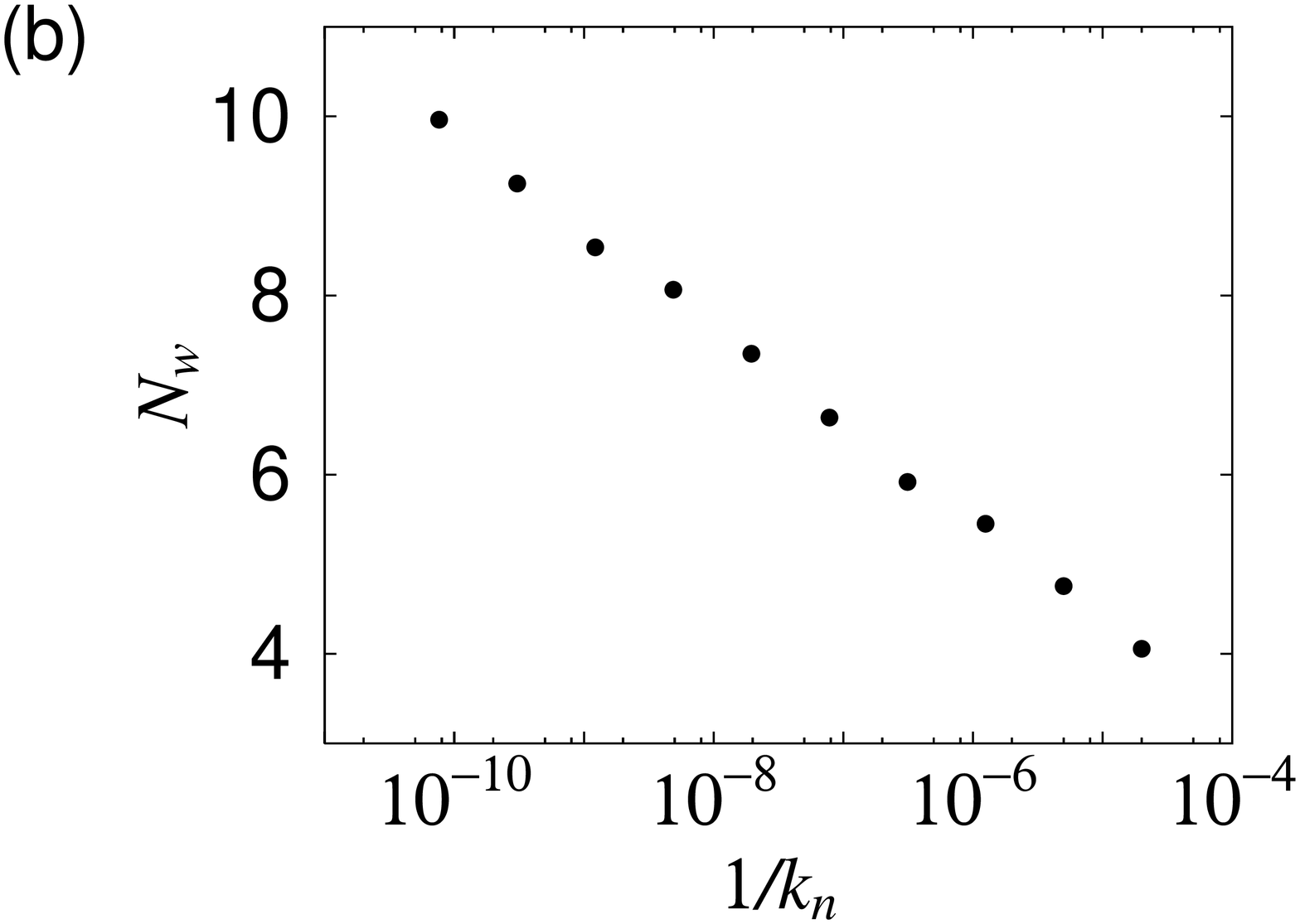}
\includegraphics[angle=-90,width=7cm]{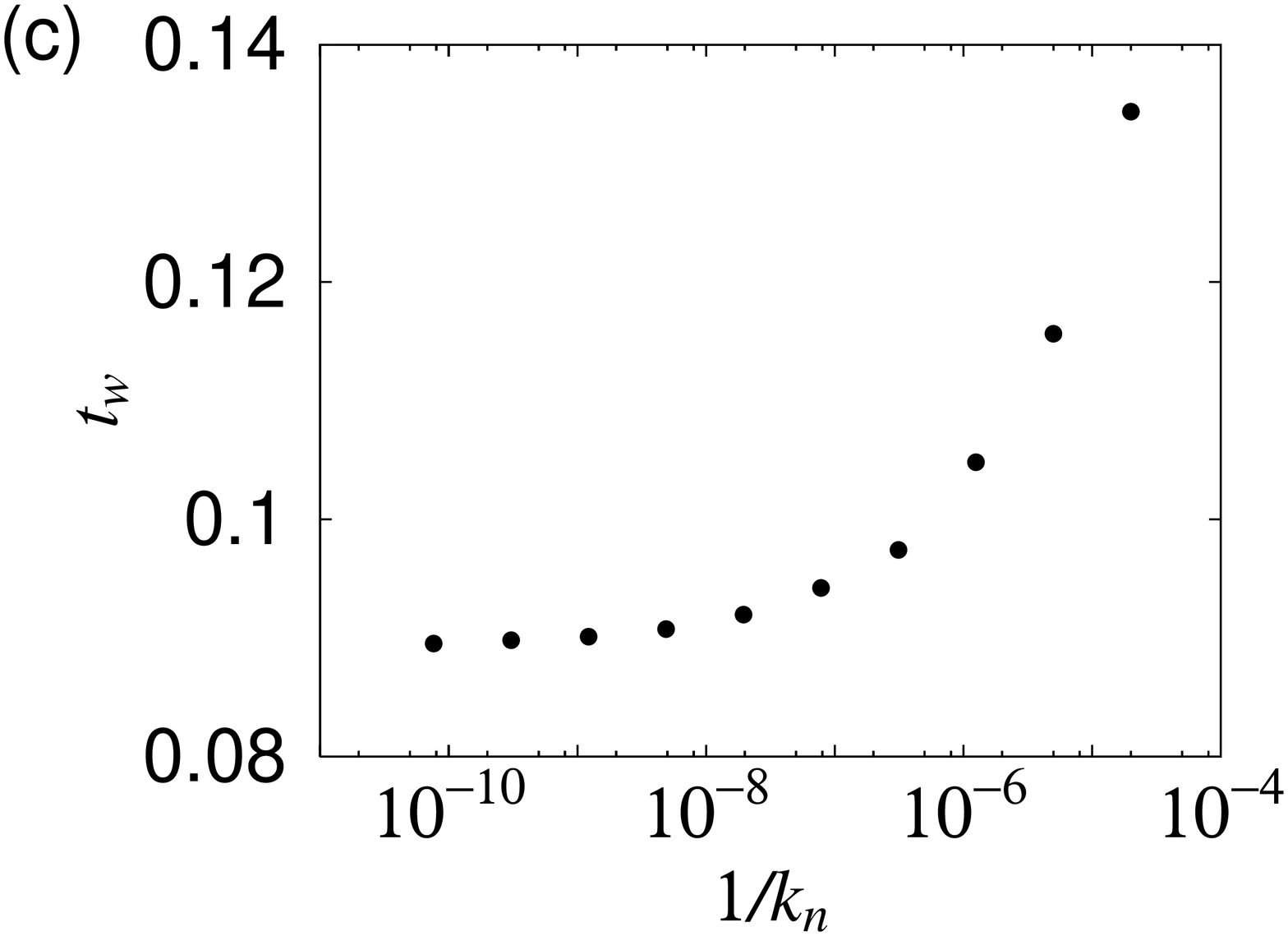}
\end{center}
\caption{The stiffness dependence of
(a) the time averaged kinetic energy of the particle $E$,
(b) the collision rate between the slope and 
the particle $N_w$,
and (c) the contact time fraction between the slope and 
the particle $t_w$.}
\label{stiff1}
\end{figure}

In Fig. \ref{stiff1} (a),
the $k_n$ dependence of the time averaged kinetic energy $E$ is shown.
It is shown that the energy is an increasing function of $1/k_n$
in the softer region ($1/k_n \gtrsim 10^{-7}$),
but no systematic $1/k_n$ dependence of $E$ exists
for $1/k_n\lesssim 10^{-7}$.
The average collision rate (number of collisions per unit time)
between the slope and the particle
, $N_w$, shows logarithmic dependence
on $1/k_n$ in the whole region (Fig. \ref{stiff1} (b)).
From Fig. \ref{stiff1} (c), we also find that 
the average contact time fraction with the slope, $t_w$, 
is a decreasing function of $k_n$ in the soft region, 
but it seems to approach  
a constant value for large enough $k_n$.

The logarithmic $k_n$ dependences of $N_w$ and 
the constant $t_w$ in large $k_n$ region
agree with our previous analysis of
``inelastic collapse under gravity''
in the soft sphere model in Section \ref{sec:2.4}.
The motion of the particle in one period is
as follows;
when the particle comes to a bump (a particle attached to the slope),
the particle jumps up, bounces on the bump many times,
loses the relative velocity, and finally rolls down
keeping in contact with the bump.
Therefore, 
the contact time fraction has a finite value
even in the hard sphere limit
due to the rolling motion at the last part.
$N_w$ increases logarithmically in the hard sphere limit
as has been expected from Eq. (\ref{eq:logk}).

\subsection{Collisional flow}
\begin{figure}[thb]
\begin{center}
\includegraphics[angle=-90,width=7cm,angle=-26.7]{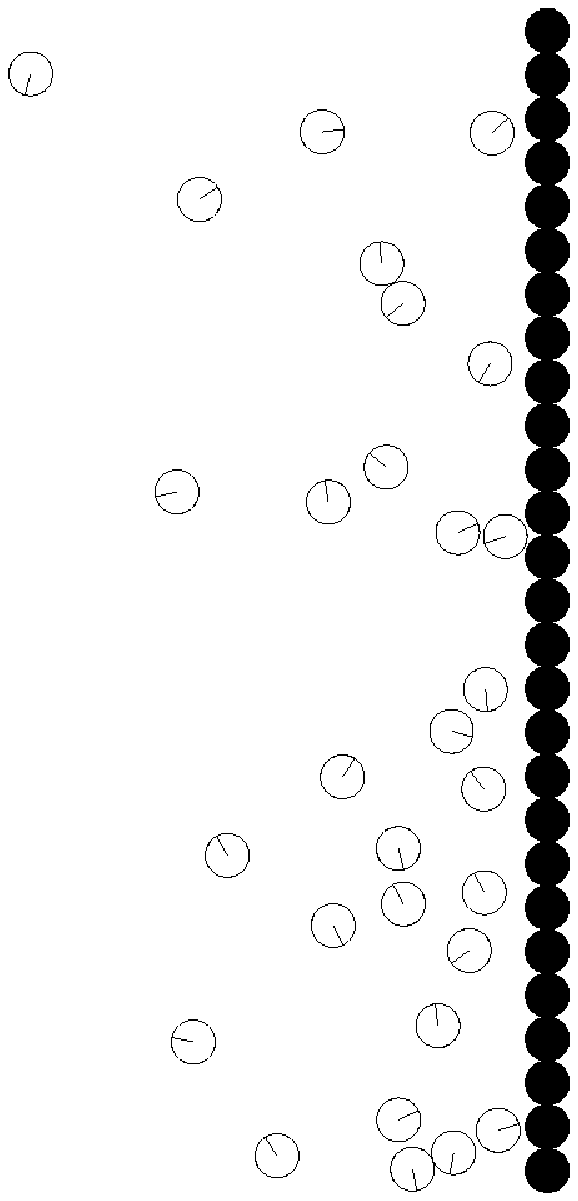}
\end{center}
\caption{A snapshot of the dilute collisional flow.}
\label{snapcol}
\end{figure}

Next we consider the steady state of the collisional flow.
The system considered is shown in Fig. \ref{snapcol}. 
The particles are monodisperse, and
the slope is made rough as in the single particle case.
The periodic boundary condition is
adopted in the flow direction ($x$ direction).
The length of the slope is $L=50.1$ and the number of the 
particles attached to the slope is $50$.
The number of flowing particles is also $50$,
namely, the number of the particles per unit length along the slope
is about $1$.
The inclination angle is set to be $\sin\theta=0.45$.
The initial configuration of particles is the row of
50 particles at rest with regular spacing
in the $x$-direction, but each particle is
at random height in the $y$-direction.
After a short initial transient,
if the total kinetic energy fluctuates around a certain value,
we consider it as the steady flow.
All the data were taken in this regime and averaged over 
the time period of $1500$.

\begin{figure}[thb]
\begin{center}
\includegraphics[angle=-90,width=7cm]{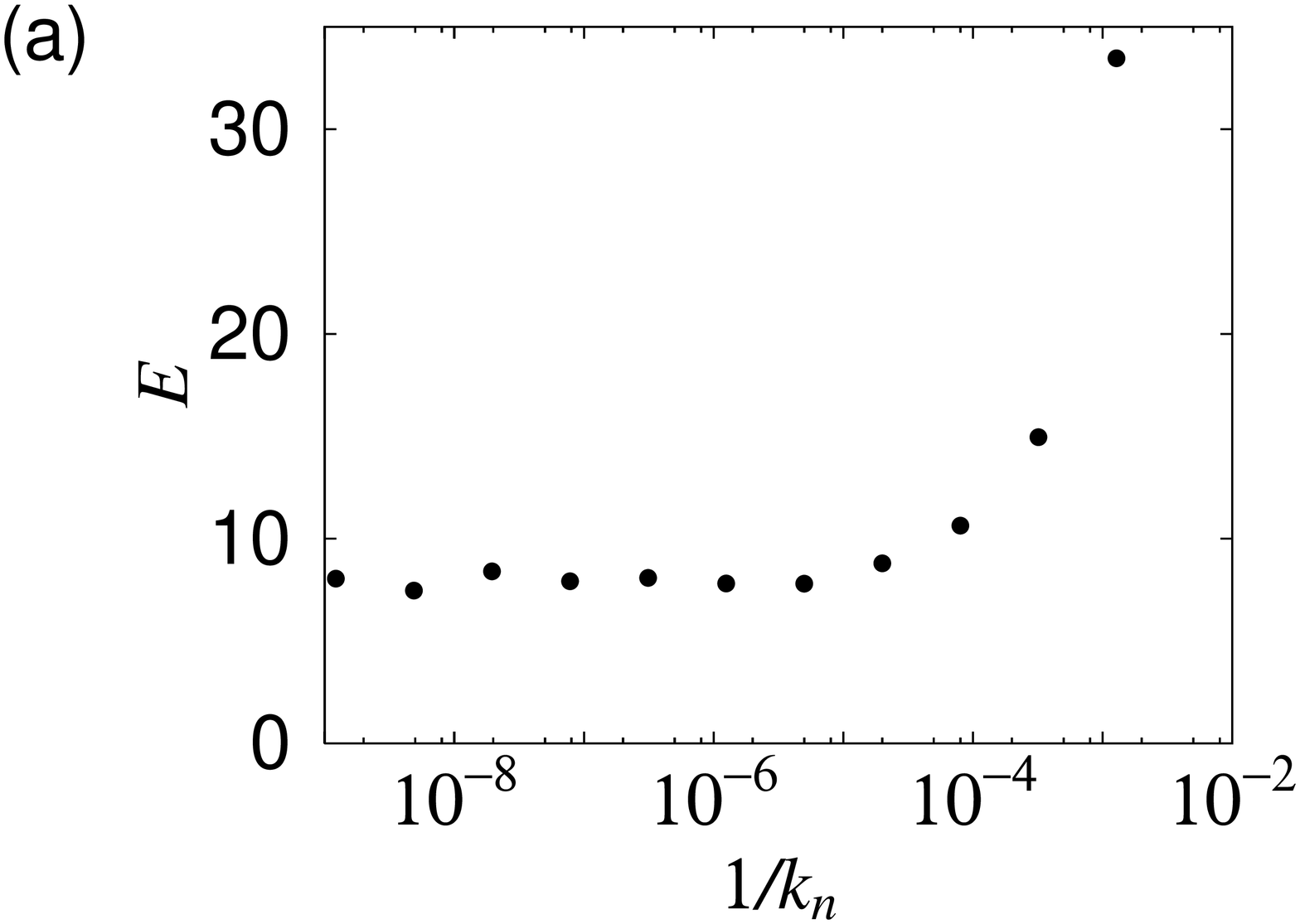}
\includegraphics[angle=-90,width=7cm]{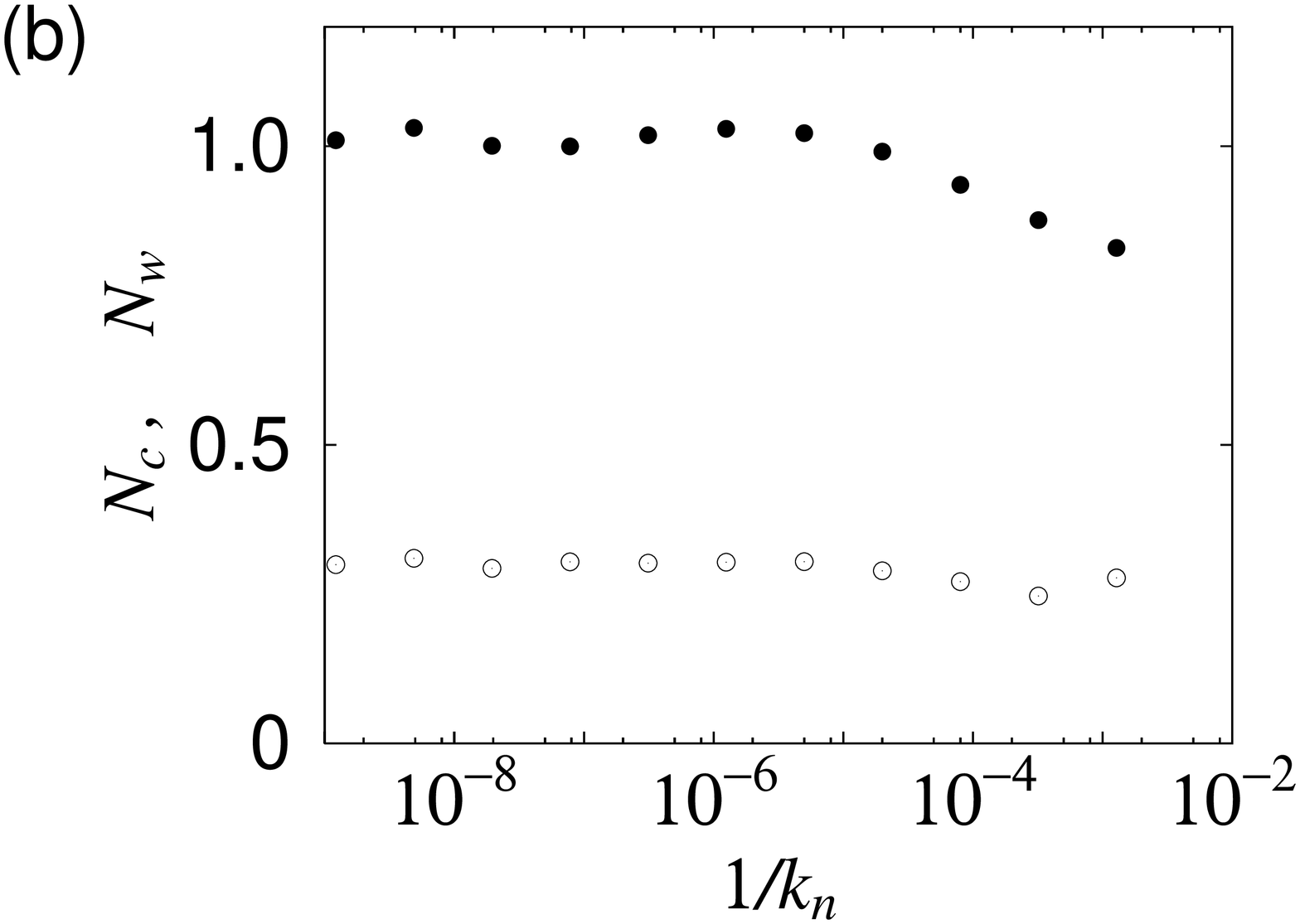}
\includegraphics[angle=-90,width=7cm]{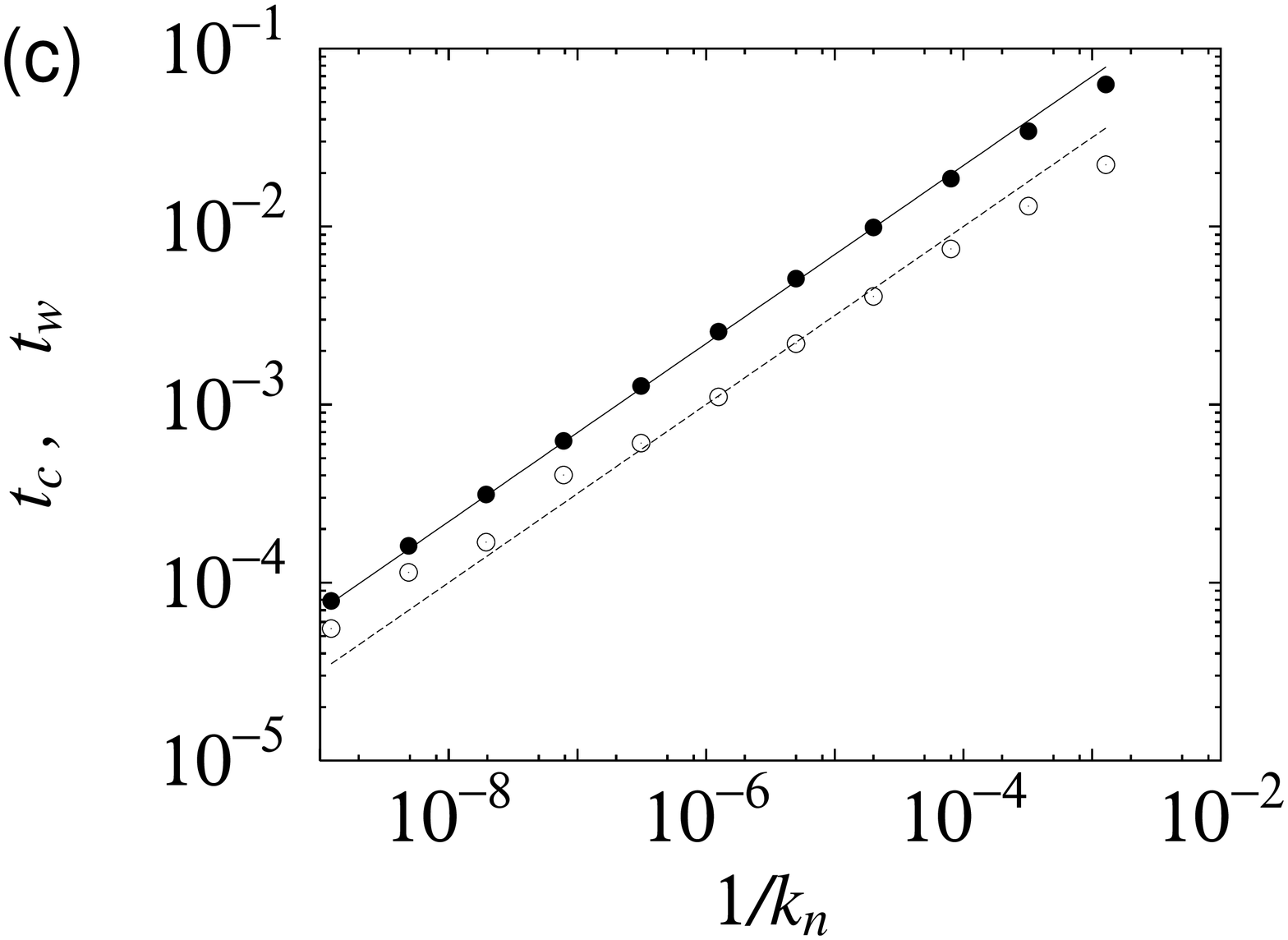}
\includegraphics[angle=-90,width=7cm]{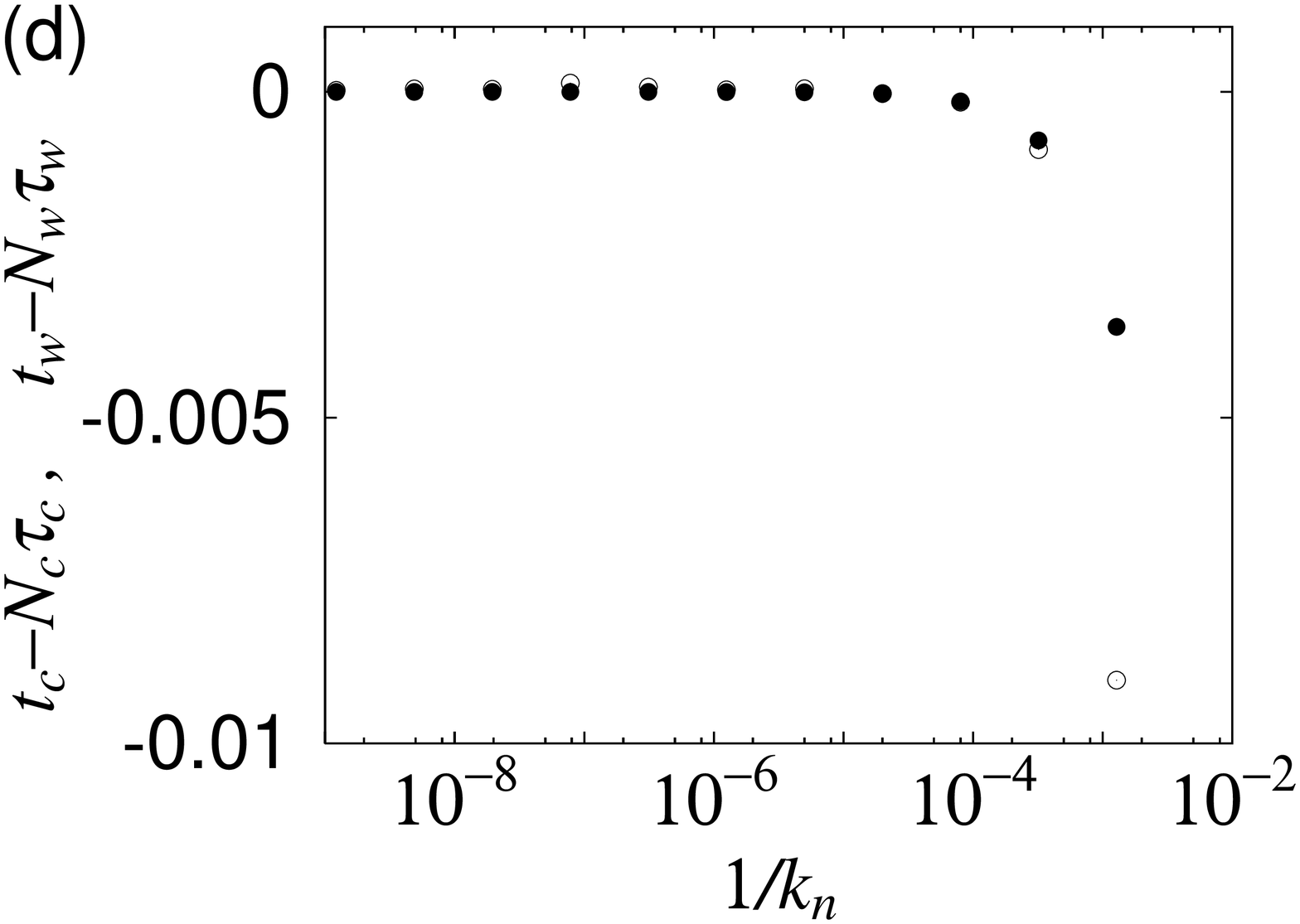}
\end{center}
\caption{The stiffness dependences of
(a) the averaged kinetic energy per one particle $E$,
(b) the averaged collision rates between particles $N_c$ (filled circles)
and between particles and the floor $N_w$ (open circles),
(c) the averaged contact time fractions 
between particles $t_c$ (filled circles)
and between particles and the floor $t_w$ (open circles),
(d) the estimated multiple contact time fractions,
$t_c-N_c \tau_c$ (filled circles) and $t_w-N_w \tau_w$
(open circles).
The solid and the dashed lines in (c) are proportional to
 $1/\sqrt{k_n}$.}
\label{col}
\end{figure}

\begin{figure}[thb]
\begin{center}
\includegraphics[angle=-90,width=3.5cm]{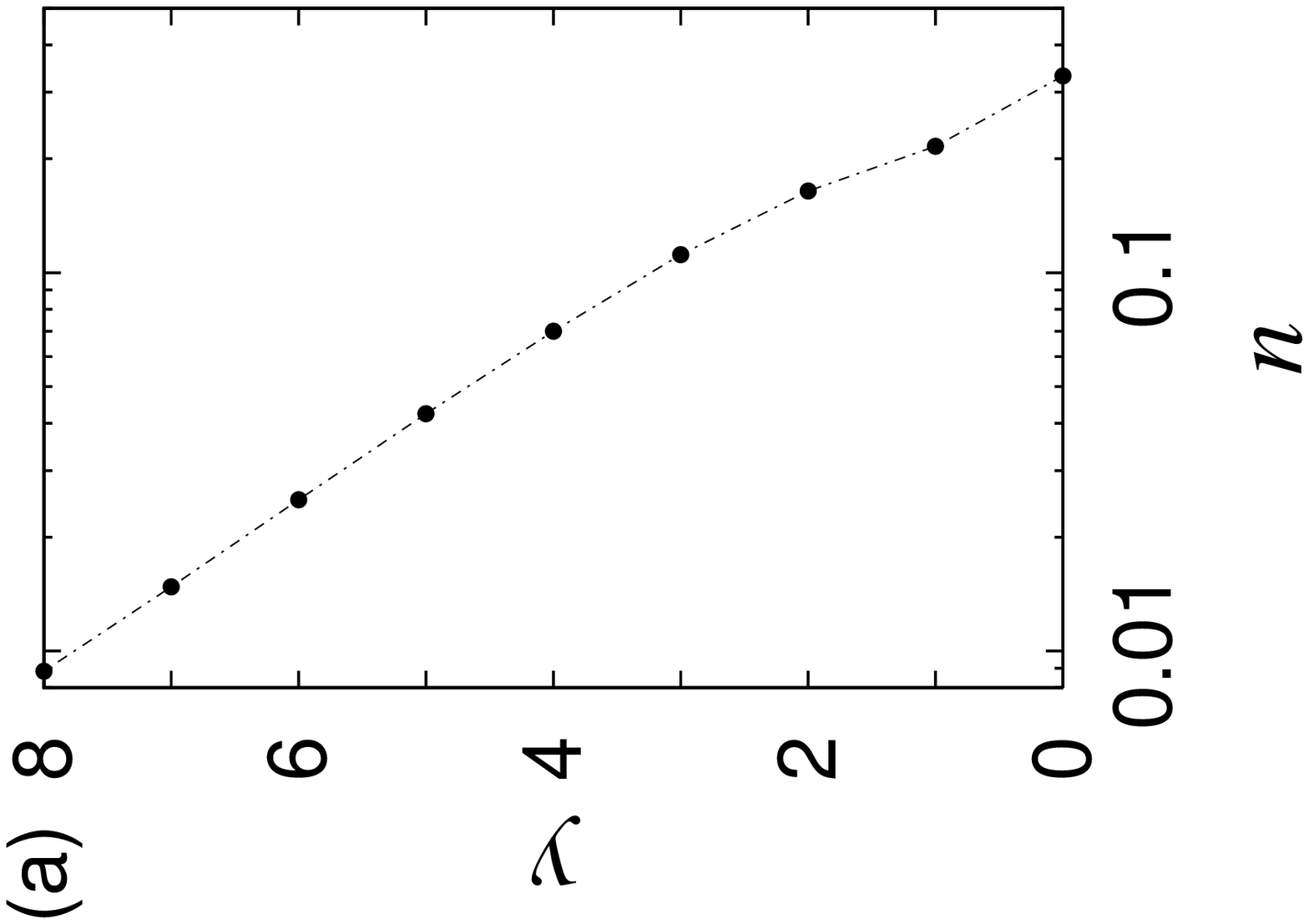}
\includegraphics[angle=-90,width=3.5cm]{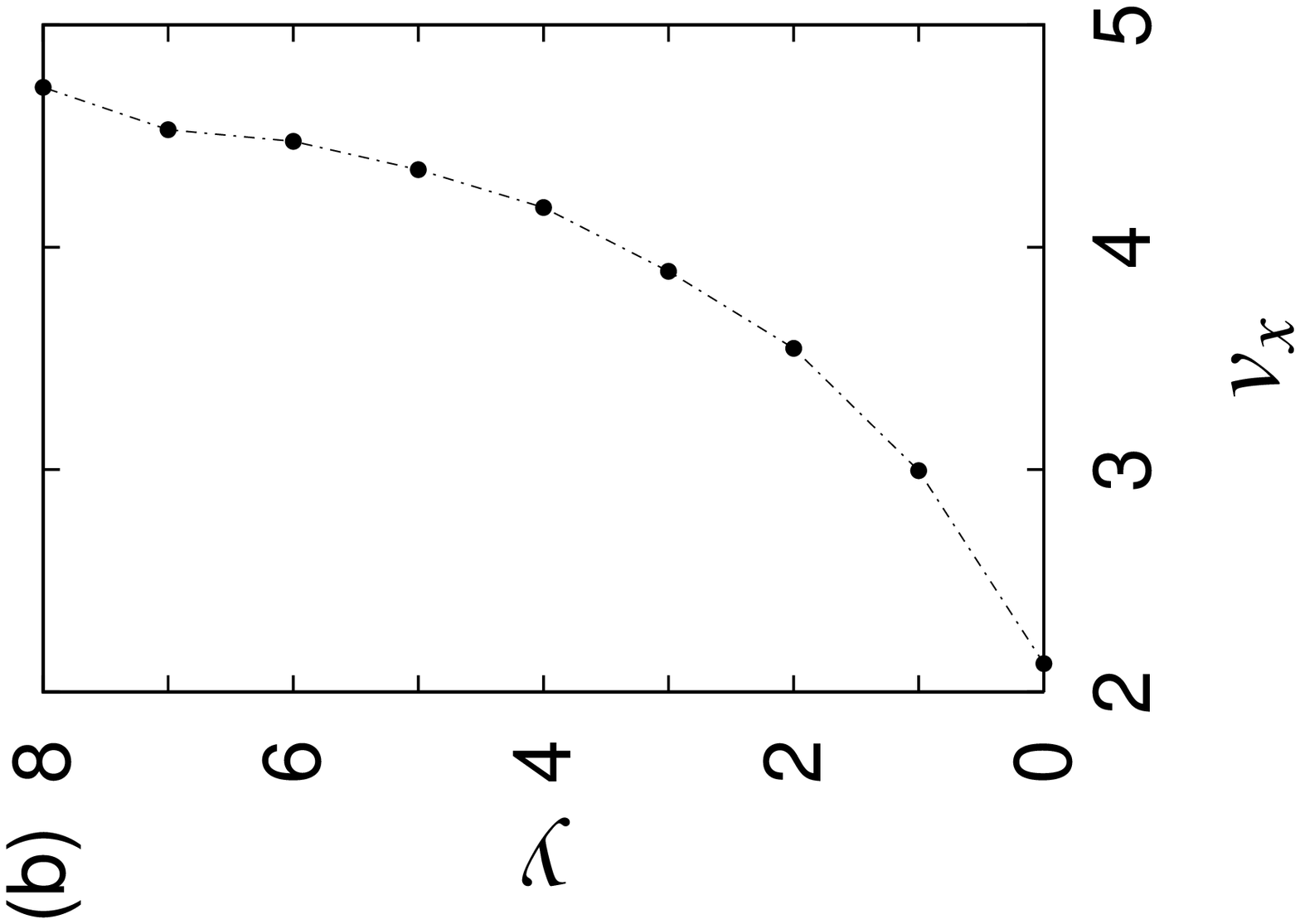}
\end{center}
\caption{The number density profile (a) and the velocity profile (b) 
of the collisional flow.}
\label{profcol}
\end{figure}

As can be seen in the snapshot, Fig. \ref{snapcol},
the particles bounce and the number of particles in contact
is very small. 
In Fig. \ref{col} (a), the averaged 
kinetic energy per one particle, $E$, is shown.
Though $E$ becomes larger as the particles become softer 
in $1/k_n\gtrsim 10^{-5}$,
the systematic dependence of $E$ on $k_n$ disappears
for large enough $k_n$ ($1/k_n\lesssim 10^{-6}$) \cite{commentyuu}.
The $y$-dependence of the average number density and the flow speed 
in this region
are also shown in Fig. \ref{profcol} (a) and (b).

In Fig. \ref{col} (b), the average collision rates 
between particles, $N_c$ (filled circles), and
between particles and the slope, $N_w$ (open circles),
per particle are plotted vs $1/k_n$.
Both of them stay roughly constant
in the region where $E$ is almost constant,
$1/k_n\lesssim 10^{-6}$.
On the other hand, the average time fractions during which 
a particle is in contact with other particles, $t_c$ 
(filled circles), and
in contact with the slope, $t_w$ (open circles), 
decrease systematically as $k_n$ increases as shown in Fig. \ref{col} (c).
Both the solid and the dashed lines are proportional to $1/\sqrt{k_n}$,
namely, $t_c$ decreases in the 
same manner as $\tau_c$ in Eq. (\ref{eq:tauc2}).
Actually, the collision time fractions $t_c$ 
and $t_w$ converge to $N_c \tau_c$ and
$N_w \tau_w$ ($\tau_w$ is the duration of 
a normal collision of a particle and the floor \cite{comment}), respectively,
in the hard sphere limit.
The differences $t_c-N_c \tau_c$ 
and $t_w-N_w \tau_w$ are plotted 
in Fig. \ref{col} (d) to 
show that they go to zero very rapidly upon increasing $k_n$.
This means that the interactions of the soft sphere model
in the collisional flow regime 
converge to those of the inelastic hard sphere model
with binary collisions.

If we look carefully,
however, in the large $k_n$ region in Fig. \ref{col} (c),
we can see slight deviation of $t_w$ from the dashed 
lines; it decreases slower than $1/\sqrt{k_n}$.
This tendency may indicate that 
$t_w$ remains finite in the hard sphere limit:
Actually, in the event-driven simulation of the hard sphere model,
we always found the inelastic collapse as long 
as the restitution constant between a particle and the floor
is less than $1$.
This suggests that $N_w$ should diverge 
and $t_w$ should remain finite
in the hard sphere limit because of 
the inelastic collapse due to gravity.
The slight deviation of $t_w$ from the dashed line
in Fig. \ref{col} (c) may be a symptom of it,
while we cannot see the logarithmic divergence 
in $N_w$.

\subsection{Frictional flow}
Now we study the steady state of the dense frictional flow.
We adopted the flat boundary with slope length $L=10.02$
and imposed the periodic boundary condition in
the flow direction. 
In order to avoid crystallization,
we used the polydisperse particles 
with the uniform distribution of diameter 
from $0.8$ to $1.0$. 
The number of particles in the system is $100$.
The inclination angle $\theta$ 
is set to be $\sin\theta=0.20$
and the initial condition is given in the similar way with the collisional
flow. The ten rows of ten particles with regular spacing in 
the $x$-direction are at rest with large enough spacings between rows in 
the $y$-direction so that particles do not overlap;
only the particles in the top row are at random heights in $y$-direction.

\begin{figure}[thb]
\begin{center}
\includegraphics[angle=-90,width=7cm,angle=-11.5]{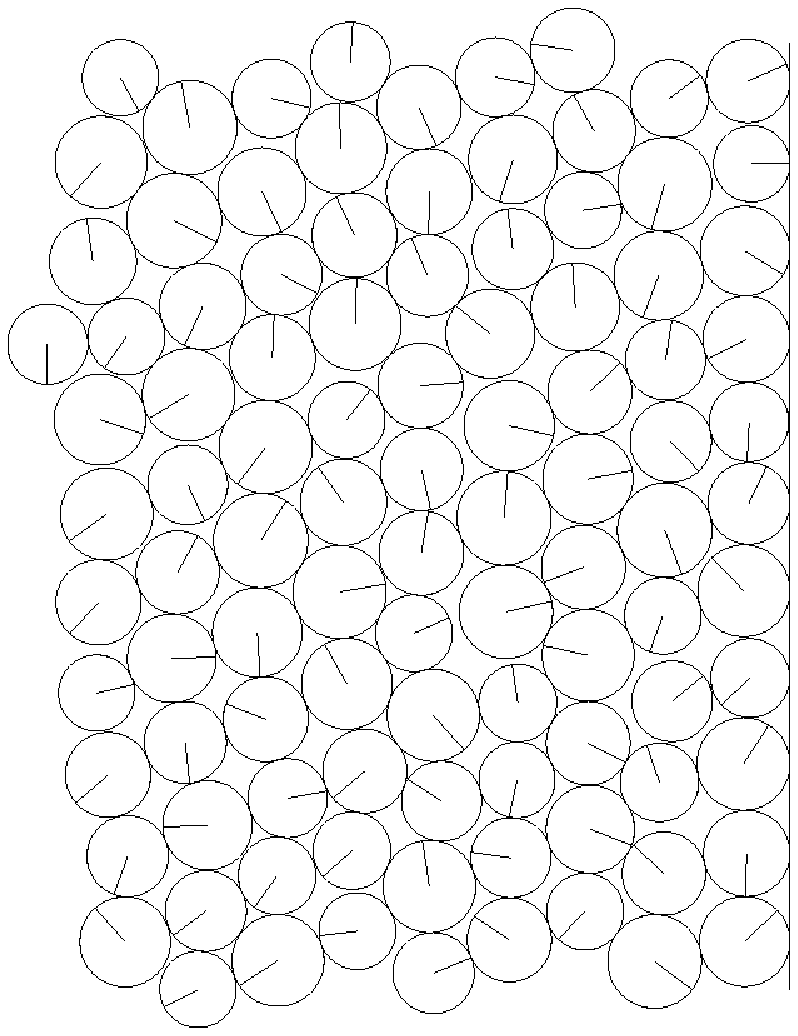}
\end{center}
\caption{A snapshot of the dense frictional flow.}
\label{snapfri}
\end{figure}

\begin{figure}[thb]
\begin{center}
\includegraphics[angle=-90,width=7cm]{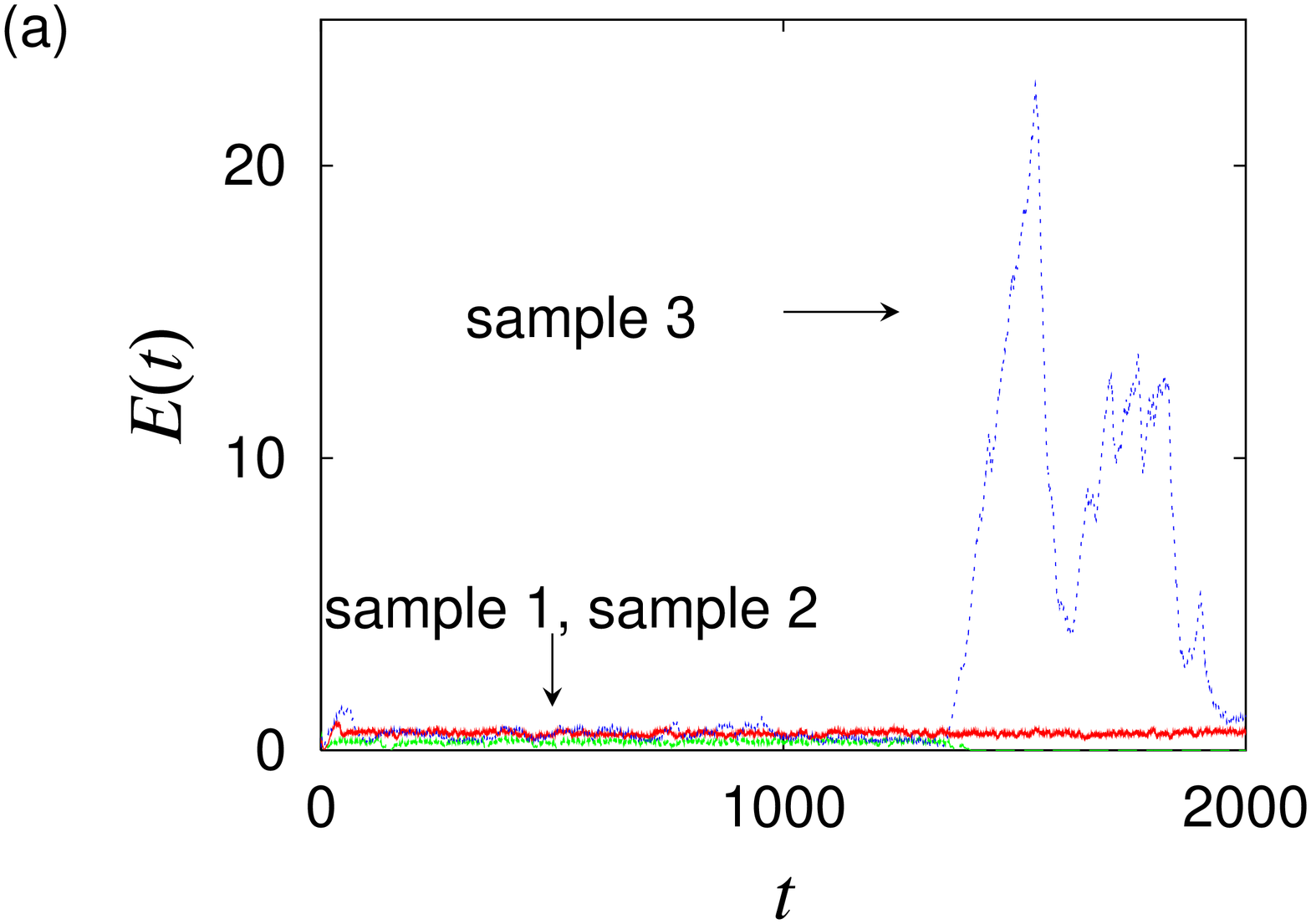}
\includegraphics[angle=-90,width=7cm]{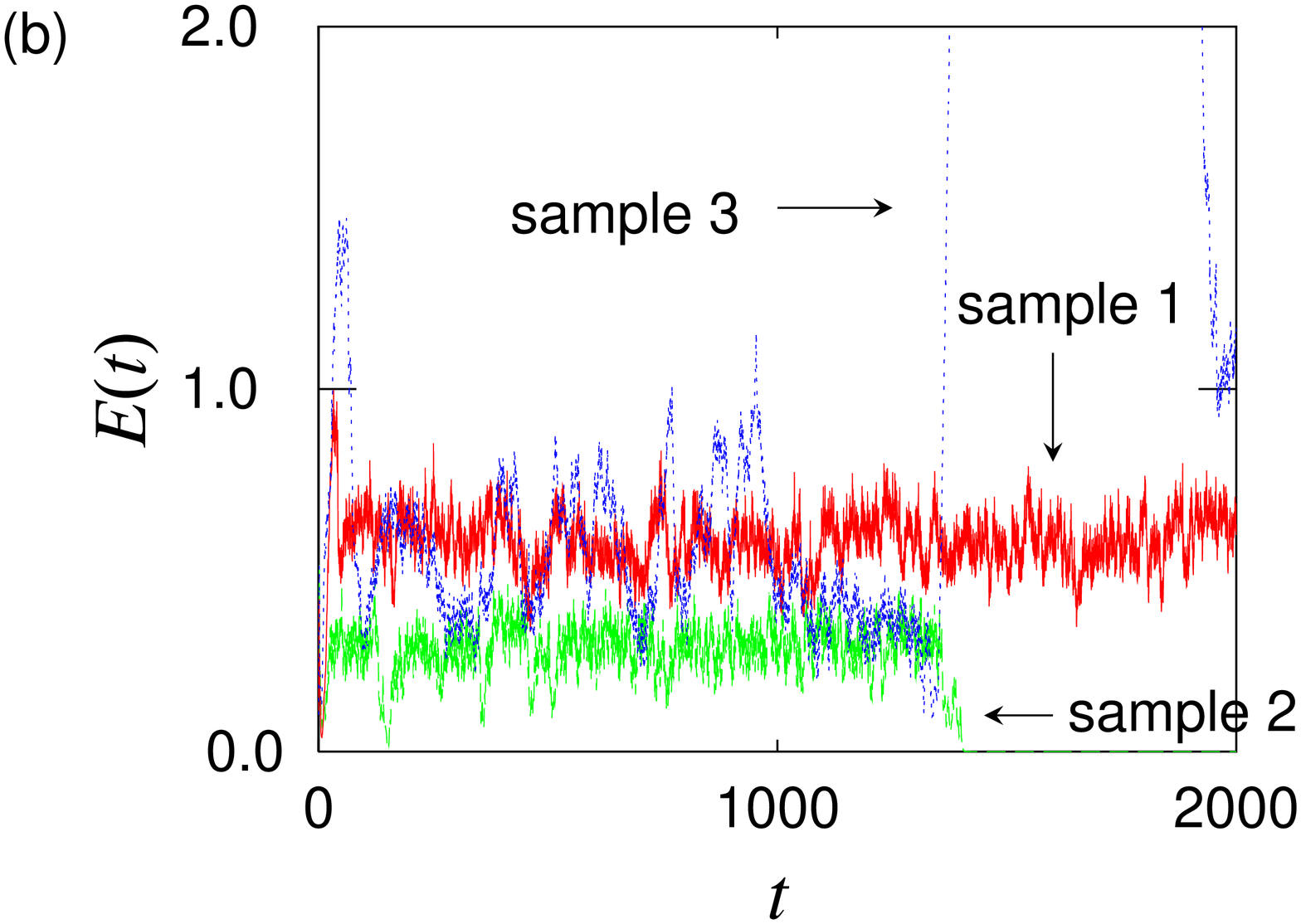}
\end{center}
\caption{Time evolution of the kinetic energy per particle
of three samples (sample 1: red solid line, sample 2: green
dashed line, sample 3: blue dotted line).
(b) is the magnification of (a).
Sample 1 with $k_n=2^{-1}\times 10^5$ shows steady behavior 
within the threshold.
Sample 2 and 3 are with $k_n=2^{-7}\times 10^5$.
Sample 2 shows steady behavior for a while
but finally stops.
$E(t)$ of sample 3 shoots up suddenly 
at $t\sim 1300$
when one of the particles in the bottom layer runs on other particles.
}
\label{timee}
\end{figure}

It turns out that 
the steady state is not unique and fluctuation is large.
Depending on the initial conditions, 
the particles may flow with a different value of average kinetic energy,
or in some cases, the whole system stops. 
We presume that this is mainly because the system is not large enough:
Grains form a layer structure as in Fig. \ref{snapfri},
and the flow velocity strongly depends on
the configuration of particles in the bottom layer.
The time evolutions of the kinetic energy per
particle, $E(t)$, of three samples with different 
configurations in the bottom layer are shown in Fig \ref{timee}.
Sample 1 shows stable behavior while 
sample 2 eventually stops after running with lower energy for
some time. Sample 3 shows larger fluctuation with higher energy;
the sudden change of $E(t)$ in sample 3 
results from the change in the configuration of the bottom layer.
This large fluctuation should be averaged out
if we could simulate large enough system for long enough time,
but the amount of computation is too large
especially for the system with stiff particles.

In order to make meaningful comparison out of these largely
fluctuating data with a variety of behaviors,
we select simulation sequences that come 
from similar flowing behaviors in the following way.
First, we define the steady part of the time sequence in each of the
samples as the part where the width of the energy fluctuations 
is smaller than 0.45 over the time
period longer than 500.
Second, we exclude the data whose averaged energy is out of
the range [0.4,0.8] \cite{select}. 

Many of the excluded data by this criterion 
show quite different flowing behaviors.
We use only the data selected from this criterion to calculate time
averages of physical quantities.

\begin{figure}[thb]
\begin{center}
\includegraphics[angle=-90,width=3.5cm]{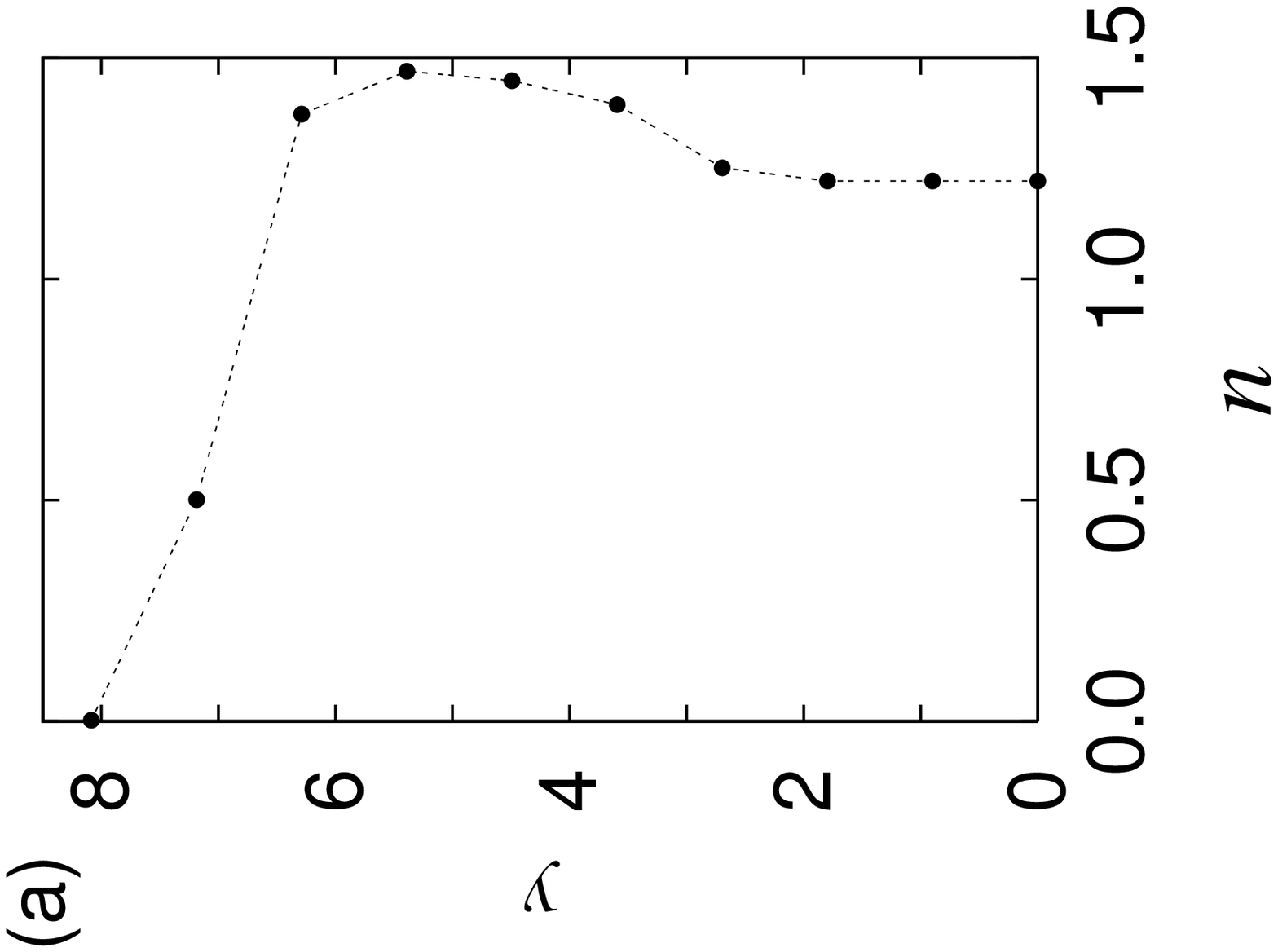}
\includegraphics[angle=-90,width=3.5cm]{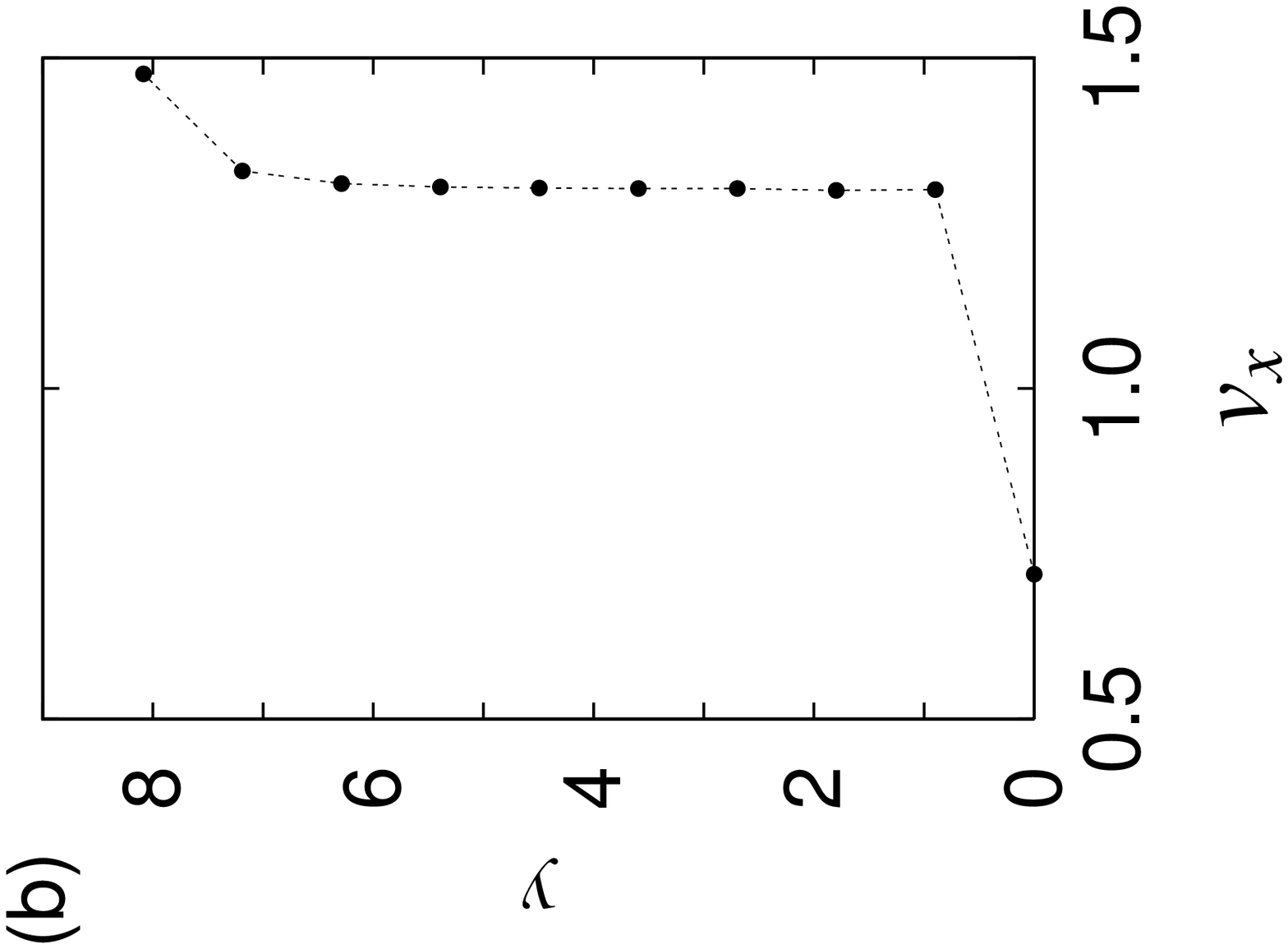}
\end{center}
\caption{The density (a) and velocity (b) profiles of the frictional flow.}
\label{proffri}
\end{figure}

\begin{figure}[thb]
\begin{center}
\includegraphics[angle=-90,width=7cm]{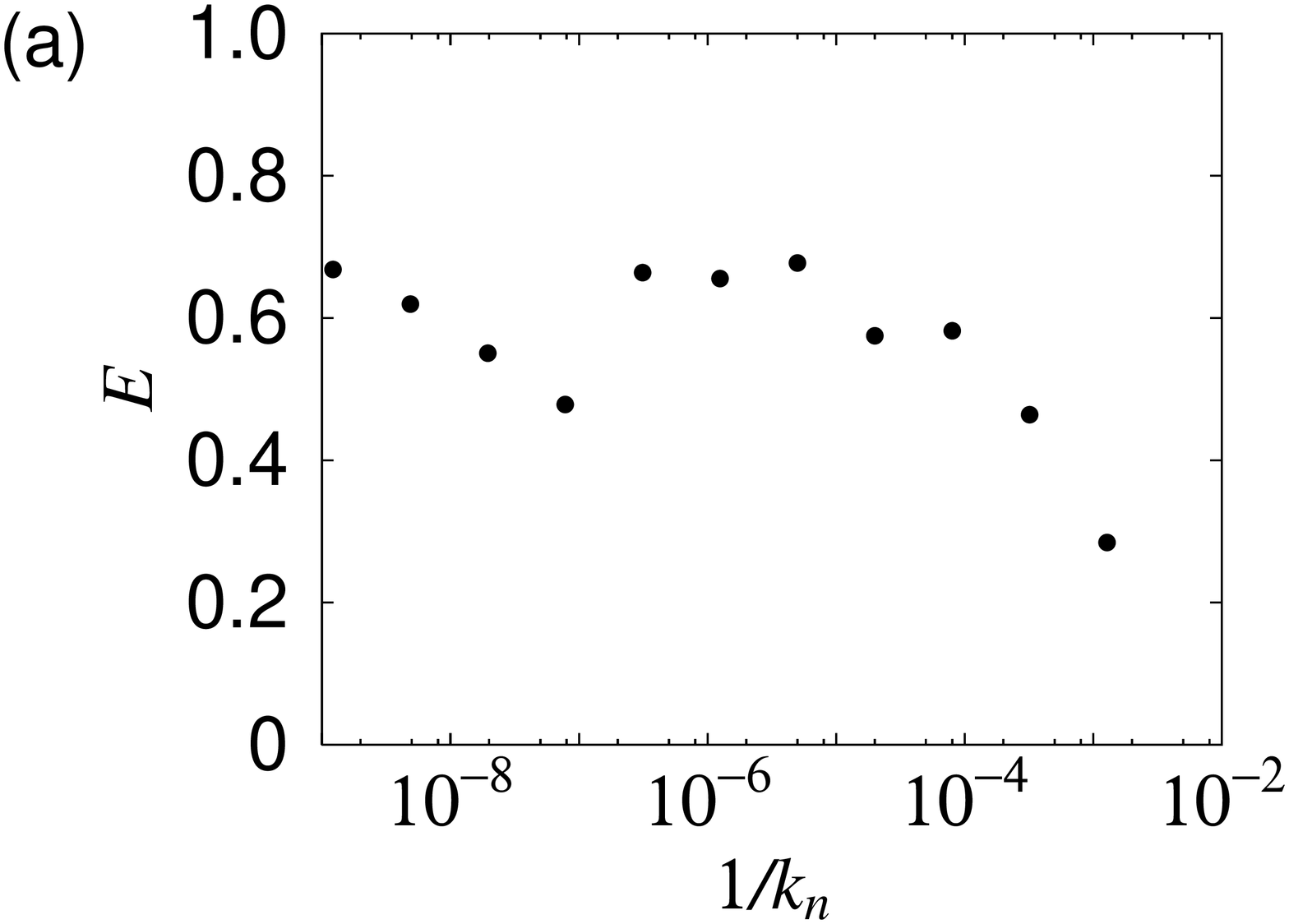}
\includegraphics[angle=-90,width=7cm]{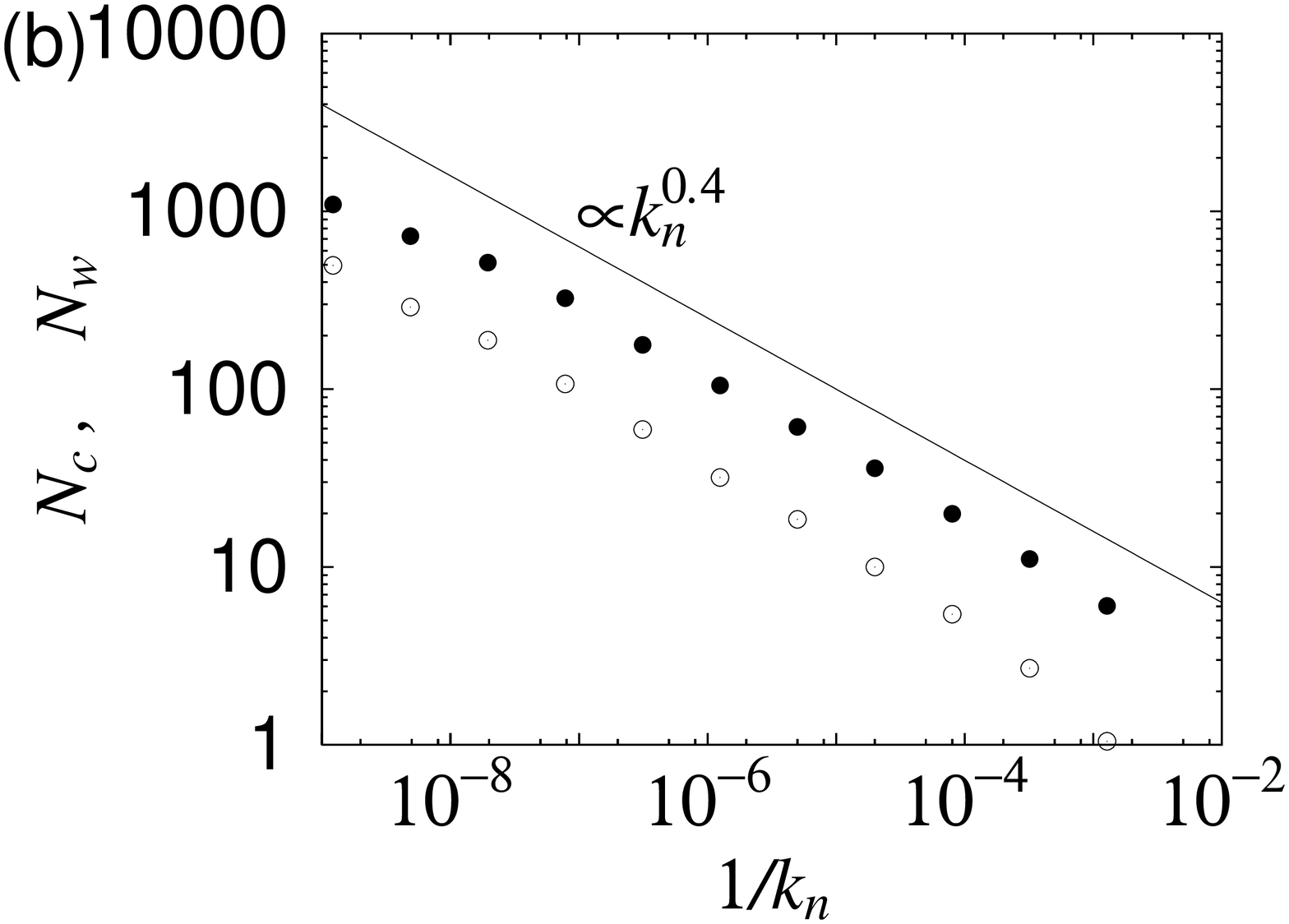}
\includegraphics[angle=-90,width=7cm]{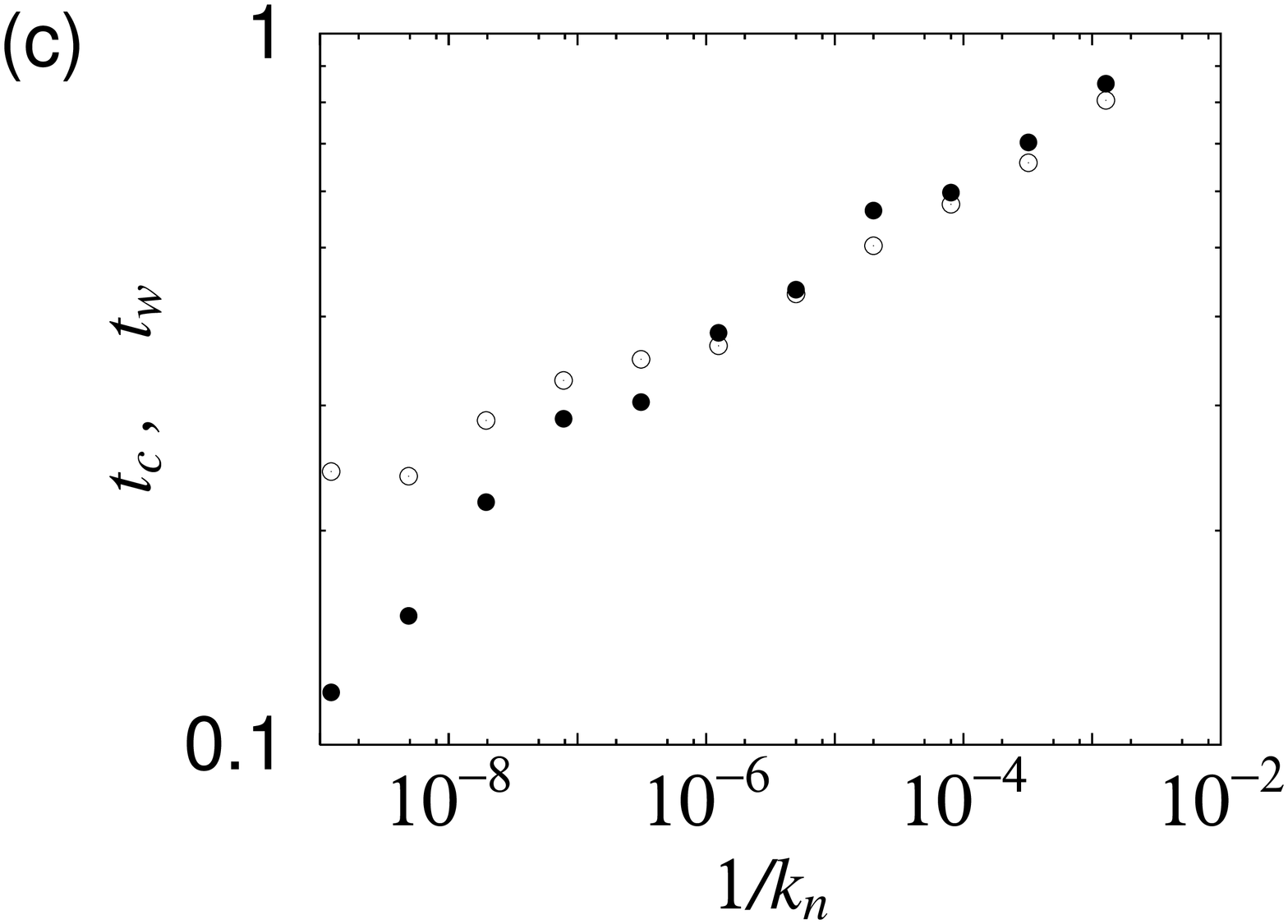}
\includegraphics[angle=-90,width=7cm]{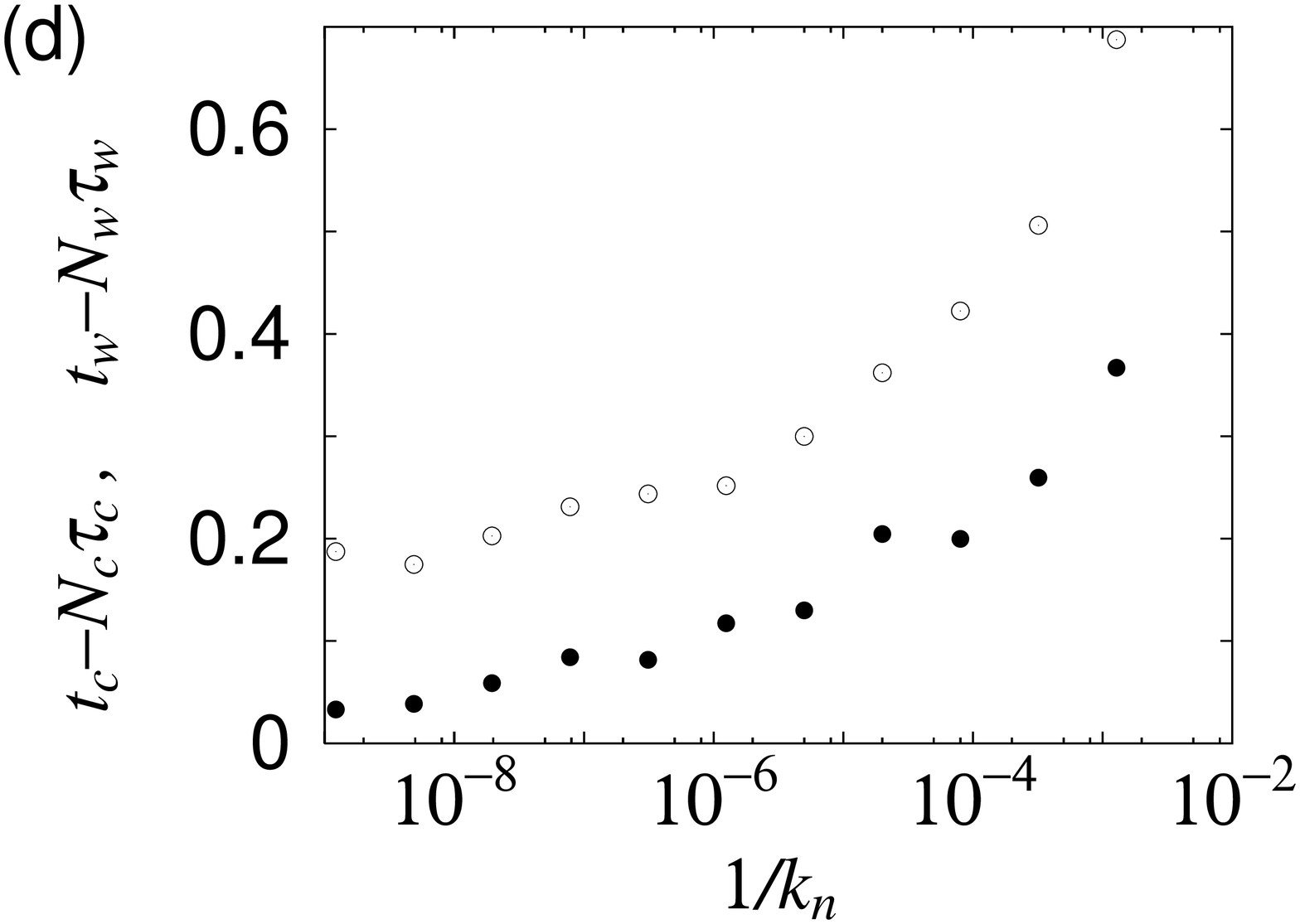}
\end{center}
\caption{ The stiffness dependences of
(a) the averaged kinetic energy $E$,
(b) the averaged collision rates between particles $N_c$ (filled circles)
and between particles and the floor $N_w$ (open circles),
(c) the averaged contact time fractions 
between particles $t_c$ (filled circles)
and between particles and the floor $t_w$ (open circles),
and (d) the estimated multiple contact time fractions,
$t_c-N_c \tau_c$ (filled circles) 
and $t_w-N_w \tau_w$ (open circles).
In (b) the solid line proportional to $k_n^{0.4}$ 
is shown for the guide of the eyes.
}\label{fri}
\end{figure}

In the snapshot of the frictional flow, 
Fig. \ref{snapfri}, most of the particles
seem to be in contact with each other and form a layer structure.
The $y$-dependence of the average density and the flow speed are
shown in Figs. \ref{proffri} (a) and (b), respectively.
We can see that the relative motion between layers 
is largest at the bottom and very small in the bulk.
The stiffness dependence of 
the average kinetic energy $E$ is shown in Fig. \ref{fri} (a);
the data are scattered due to the non-uniqueness of the steady state.

Nevertheless, the average collision rates
show systematic dependence as shown in Fig. \ref{fri} (b).
Here, the definition of $N_c$ ($t_c$) is the same
as that for the collisional flow,
namely, it is the average collision rate (contact time fraction) 
between particles {\it per particle in the system}.
On the other hand,
$N_w$ ($t_w$) is defined differently;
it is the collision rate (the 
contact time fraction) between the particles and the slope 
{\it per particle in the bottom layer}, 
because other particles never touch the slope.

In Fig. \ref{fri} (b), we can see that 
$N_c$ (filled circles) and $N_w$ (open circles)
increase very rapidly as $k_n$ becomes larger:
they diverge as a power of $k_n$.
This is quite different from the behavior in the collisional flow
in which they are almost constant.
Furthermore, this increase is faster than the logarithmic divergence 
found in the single particle case 
(see Fig. \ref{stiff1} (b)).

The contact time fractions, $t_c$ (filled circles) and 
$t_w$ (open circles), 
decrease as shown in Fig. \ref{fri} (c).
The main reason why $t_c$ and $t_w$ 
decrease is that
$\tau_c$ and $\tau_w$ decrease faster than $N_c$ and $N_w$,
namely, the contact time fractions 
estimated from the duration of a binary collision,
$N_c \tau_c$ and $N_w \tau_w$, continue to decrease.
Actually, the decrease of the contact time fraction and 
the increase of the collision rate are natural because a longer multiple
contact breaks up into shorter binary contacts as the particles become
stiffer. 

These contact time fractions, $t_c$ and $t_w$, however,
do {\it not} converge to $N_c  \tau_c$ and $N_w \tau_w$,
respectively. As shown in Fig. \ref{fri} (d),
$t_c-N_c \tau_c$ (filled circles) and 
$t_w-N_w \tau_w$ (open circles)
are very large as compared to those in the collisional flow,
even in the stiffest region. 
The comparison of this with the rapid convergence in the 
collisional flow regime (Fig. \ref{col} (d))
indicates that there remains finite multiple contact 
time in the hard sphere limit and the interaction
in the frictional flow can never be
considered as many, or even infinite, instantaneous binary collisions.
The particles experience the lasting 
multiple contact even in the hard sphere limit. 

\section{Summary and Discussion}
\label{sec:4}
The inelastic hard sphere limit of
granular flow has been investigated numerically 
in the steady states of
(i) a single particle rolling down the slope,
(ii) the dilute collisional flow,
and (iii) the dense frictional flow.
In (i), it has been found that 
the ``inelastic collapse'' between
the particle and the slope occurs in the
hard sphere limit due to gravity, and 
the contact time fraction between the particle
and the slope remains finite.
In (ii), the collision rates $N_c$ and $N_w$
are almost constant when particles are stiff enough.
The contact time fraction between particles $t_c$
approaches zero as $k_n$ increase
in the same manner with the duration of contact
for binary collision, $\tau_c$.
This means that the interaction between 
particles in the hard sphere limit 
can be expressed by binary collisions in the inelastic hard sphere
model. On the other hand, the decrease in $t_w$ 
is slightly slower than $1/\sqrt{k_n}$ in the harder region,
which can be a sign of the inelastic collapse 
between a particle and the slope.

In the case of the frictional flow (iii), the situation is not simple:
Although the contact time fractions $t_c$ and $t_w$
decrease upon increasing $k_n$,
the collision rates $N_c$ and $N_w$ increase
as a power of $k_n$,
which is faster than the logarithmic divergence found in the 
single particle case (i).
The origin of this power divergence of collision rate
does not seem to be simple
because it is a property of the steady state,
not a particular dynamical trajectory of the system.

The multiple contact time fraction may be estimated by
$t_c-N_c \tau_c$ and $t_w-N_w \tau_w$.
They were found to be quite large
compared to those for the collisional flow, and seems to
remain finite even in the infinite stiffness limit:
This suggests
that the interaction in the frictional
flow can never be considered as infinite number of 
binary collisions. Even in the hard sphere limit,
particles experience the lasting multiple contact.

The non-negligible fraction of multiple contact
in the hard sphere limit
implies the existence of the network of contacting grains
even though they are flowing.
The models of dense flows should 
consider the effect of this lasting contacts.

Here we have investigated only the two cases, i.e.,
the collisional flow and the frictional flow.
The system should undergo the transition 
between the two flows
if we change continuously the parameters 
such as the inclination angle, the roughness of the slope,
or the density of particles.
It is interesting to investigate how the transition 
occurs by looking at the quantities measured in this paper,
because their stiffness dependences are qualitatively different 
in the two flows.

\begin{acknowledgments}
This research was partially supported by 
Hosokawa powder technology foundation,
the Japan Society for the Promotion of Science,
and Grant-in-Aid from the Ministry of Education, Science, 
Sports and Culture of Japan.
   
\end{acknowledgments}


\begin{thebibliography}{99}
\bibitem{comment}
When we fix one of the particles on the slope and 
employ the same force law as Eq. (\ref{eq:fn}), 
then we have the duration of contact for the normal collision 
without gravity 
$\tau_w=\pi/\sqrt{k_n-\eta_n^2/4}$ for identical particles.
The force law for the flat boundary that replace $2M$ in 
Eq. (\ref{eq:fn}) by the mass of the colliding particle $m$
gives the same duration of contact with the above one 
for the normal collision.
\bibitem{select}
As for the softest case, $1/k_n=2^{7}\times 10^{-5}$, 
we had to include the data
whose energy goes below the lower limit because of the general
tendency that the energy becomes small for the steady part of the very
soft region. For $1/k_n\lesssim 10^{-5}$, however, we did not see any
general trend in the average energy.
\end{thebibliography}

\newpage

\end{document}